\documentclass[aps,pra,showpacs,amsmath,amssymb,amsfonts,lengthcheck,longbibliography,superscriptaddress]{revtex4-2}

\usepackage{graphicx}
\usepackage[caption=false]{subfig}
\usepackage[colorlinks=true,citecolor=blue,linkcolor=blue,urlcolor=blue]{hyperref}

\usepackage{soul}
\usepackage[utf8]{inputenc}
\usepackage[T1]{fontenc}

\usepackage{mathtools}
\usepackage{SIunits}
\usepackage{braket}

\begin{document}

\title{Performance of optimal linear-response processes in driven Brownian motion far from equilibrium}

\author{Lucas P. Kamizaki}
\email{kamizaki@ifi.unicamp.br}
\affiliation{Instituto de F\'isica `Gleb Wataghin', Universidade Estadual de Campinas, 13083-859, Campinas, S\~{a}o Paulo, Brazil}
\affiliation{Instituto de F\'isica de S\~ao Carlos, Universidade de S\~ao Paulo, 13560-970, S\~ao Carlos, S\~{a}o Paulo, Brazil}

\author{Marcus V. S. Bonan\c{c}a}
\email{mbonanca@ifi.unicamp.br}
\affiliation{Instituto de F\'isica `Gleb Wataghin', Universidade Estadual de Campinas, 13083-859, Campinas, S\~{a}o Paulo, Brazil}

\author{S\'ergio R. Muniz}
\email{srmuniz@ifsc.usp.br}
\affiliation{Instituto de F\'isica de S\~ao Carlos, Universidade de S\~ao Paulo, 13560-970, S\~ao Carlos, S\~{a}o Paulo, Brazil}

\begin{abstract}
Considering the paradigmatic driven Brownian motion, we perform extensive numerical analysis on the performance of optimal linear-response processes far from equilibrium. We focus on the overdamped regime where exact optimal processes are known analytically, and most experiments operate. This allows us to compare the optimal processes obtained in linear response and address their relevance to experiments, using realistic parameter values from experiments with optical tweezers. Our results help assess the accuracy of perturbative methods in calculating the irreversible work for cases where an exact solution does not exist. For that, we present a performance metric comparing the approximate optimal solution to the exact one. Our main result is that optimal linear-response processes can perform surprisingly well, even far from where they were expected.
\end{abstract} 

\maketitle

%TC:ignore
%\keywords{Finite-time thermodynamics, Optimal protocols}
%TC:endignore

\section{Introduction}

Finite-time thermodynamic processes are ubiquitous. They are the way we control real-world systems and their environments \cite{chu2002control,vander2005control,koch2019control,kumar2020nature,deffner2020natphys}. However, the second law of thermodynamics states that nonequilibrium processes have an unavoidably higher cost than their quasistatic counterparts. Therefore, a major goal in science and engineering is understanding and searching for the minimal waste of resources to achieve a predetermined task. This leads to the fundamental problem of efficiency: finding the finite-time process with the minimum possible cost \cite{deffner2020thermcontrol}.

Taken in its full generality, the problem of finding optimal finite-time processes is a tough one, with few examples where exact solutions are known. A paradigmatic case where such solutions exist is the driven Brownian motion \cite{schmiedl2007optimal,gomez2008optimal}. Its relevance is manifested by the number of experiments to test several different nonequilibrium phenomena. Using colloidal particles or beads trapped by optical tweezers, driven Brownian motion has been used to address fluctuation theorems \cite{blickle2006prl,ciliberto2008epl,ciliberto2008jstatmech}, heat engines \cite{blickle2012nature,proesmans2016prx,roldan2016soft}, feedback processes \cite{toyabe2010experimental}, Maxwell's demons \cite{roldan2014natphys}, and Landauer's principle \cite{berut2012nature,jun2014prl,gavrilov2016prl}.  

It seems natural to expect that the general features of the physics of optimal finite-time processes in driven Brownian systems might shed light on more complicated systems. As mentioned, exact analytical expressions for optimal processes are known in this case \cite{schmiedl2007optimal,gomez2008optimal}. However, they present unexpected features such as jumps and sharp peaks that have been barely understood physically and represent a real challenge for experimental implementation. Such counter-intuitive characteristics have been reproduced by numerical approaches based on optimal control \cite{then2008pre,geiger2010pre,aurell2011prl}, but this has not clarified the role of unexpected jumps and peaks in the optimal processes.

At the same time, perturbative approaches have been developed to provide approximate optimal finite-time protocols. They express the energetic cost of a given finite-time process in terms of functionals of the corresponding protocol, which are specific for certain regimes. If, on the one hand, these formulations restrict the optimization problem to limited non-equilibrium regimes, on the other hand, they provide a better physical intuition about the energetic cost through the quantities appearing in the derived functionals. Among these perturbative formulations, the so-called geometric approach has attracted considerable attention in the last decade \cite{sivak2012prl,zulkowski2012pre,bonancca2014optimal,zulkowski2015pre,zulkowski2015pre2,sivak2016pre,rotskoff2016pre,rotskoff2017pre,lucero2019pre,scandi2019quantum,blaber2020jcp,abiuso2020prl,louwerse2022preprint,blaber2022preprint}. It has been applied to different non-equilibrium situations in biophysics \cite{lucero2019pre,blaber2020jcp,blaber2022preprint}, magnetic systems \cite{rotskoff2016pre,rotskoff2017pre,louwerse2022preprint}, heat engines \cite{abiuso2020prl}, and solid-state physics \cite{koning1997prb}. It also has been extended to quantum systems \cite{zulkowski2015pre2,scandi2019quantum}. In this approach, the energetic cost is written as the time integral of a Lagrangian, understood as a thermodynamic metric \cite{salamon1983prl,ruppeiner1995rmp,crooks2007prl}. The optimal finite-time processes, then, have the interpretation of being the corresponding geodesics. 

Here, we show that the perturbative approach derived in Ref.~\cite{bonancca2018minimal} performs quite well beyond its expected range of validity, while the performance of the geometric approach, despite its relevance, is generally worse in the far from equilibrium region.
Taking driven Brownian motion in the overdamped regime as a benchmark, we compare the performance of exact and approximate optimal protocols obtained from either Ref.~\cite{bonancca2018minimal} or the geometric approach and present it using realistic numbers, motivated by current experiments.  Our extensive numerical analysis shows a clear advantage of the perturbative formulation in describing the optimal energetic cost far from equilibrium within a range of experimentally relevant values of the parameters involved. Additionally, the optimal protocols of Ref.~\cite{bonancca2018minimal} clearly consist of continuous and smooth versions of the exact optimal protocols derived in Ref.~\cite{schmiedl2007optimal}. Therefore, in addition to being very attractive for experimental implementations, these approximate optimal protocols show that even smooth but fast changes at the right places of the process consist of a good (but unexpected) optimization strategy far from equilibrium. This fact supports the claim that the perturbative approaches, in contrast to optimal control numerical methods, can potentially increase our physical understanding of optimal non-equilibrium processes.

We try to keep the presentation self-contained, so the manuscript is organized as follows. In Secs.~\ref{sec:bmotion} and \ref{sec:stothermo}, we establish notations reviewing the standard theoretical description of Brownian motion and its corresponding stochastic thermodynamics, respectively. In Sec.~\ref{sec:exactopt}, we re-derive the exact optimal protocols for driven Brownian motion in the overdamped regime and define the performance we test numerically. In Sec.~\ref{sec:optslow}, we present the basic elements of the geometric approach applied to driven Brownian motion and derive the corresponding optimal protocols, which we use later in our performance analysis. In Sec.~\ref{sec:fastopt}, we obtain fast but weak optimal processes using the approach of Ref.~\cite{bonancca2018minimal} and obtain their performance far from equilibrium. We give our final remarks in Secs.~\ref{sec:power} and \ref{sec:conclusion}.

\section{Description of Brownian motion \label{sec:bmotion}}

\subsection{Langevin equation}

If $x(t)$ is the position at time $t$ of a small particle of mass $m$ subjected to a potential  $V(x)$ and immersed  in a liquid with friction coefficient $\gamma$ at temperature $T$, the equation describing the particle's motion is given by 
\begin{equation}
    m \frac{d^2 x(t)}{dt^2} = - \frac{d}{dx}V(x) - \gamma \frac{d x(t)}{dt} + \sqrt{2 k_B T \gamma} \xi(t),
    \label{eq: langevin}
\end{equation}
which is known as the Langevin equation. The last term on the right-hand side, $\xi(t)$, is the stochastic term. It describes a stochastic force modeled as a random Gaussian white noise with zero mean. Mathematically, this corresponds to
\begin{equation}
    \overline{\xi(t)} = 0,
    \label{Noise_mean_zero}
\end{equation}
and
\begin{equation}
    \overline{\xi(t)\xi(t^{'})} = \delta{(t-t')}.
    \label{Noise_correlation}
\end{equation}
The overbar denotes averages under different realizations of the white noise.

\subsection{Fokker-Planck equation}
Another way of describing the motion of the particle is through the time evolution of the corresponding probability distribution. In several situations, this description is more appropriate than using the Langevin equation.  Consider that $P(x,t)$ is the probability density function of finding the particle in the position $x$ at time $t$. Its time evolution is given by \cite{riskenFPE,reichl1999modern}
\begin{equation}
    \frac{\partial P(x,t)}{\partial t} = \frac{1}{\gamma}\frac{\partial}{\partial x}\left[\frac{dV(x)}{dx}P(x,t) +  k_{B} T\frac{\partial P(x,t)}{\partial x}\right]\,,
    \label{eq: Fokker-Planck}
\end{equation}
which is the Fokker-Planck equation in the strong friction limit. As usual, the probability density $P(x,t)$ satisfies the normalization condition.

\subsection{Simulations: setup and realistic parameters}

Optical tweezers are versatile tools used to trap and manipulate microscopic particles \cite{Marago2013,jones2015optical,Giesele2021}, from single atoms to macromolecules like DNA/RNA, up to living cells, and a wide range of colloidal particles \cite{kumar2020nature,ciliberto2008epl,blickle2012nature,proesmans2016prx}.

Motivated by current experiments using optical trapping and dynamical control to test and explore ideas of nonequilibrium thermodynamics \cite{otani2022,martins2021,oliveira2019}, we performed numerical simulations of a trapped spherical colloidal particle immersed in water at room temperature, using standard numerical tools \cite{volpe2013simulation}. 

For sufficiently large colloidal particles (typically with a radius, $r$, in the micrometer range) near the focus of a Gaussian laser beam, the confining optical potential near the focus can be safely approximated by a simple harmonic potential 
\begin{equation}
    V(x) = \frac{\kappa (x - x_c)^2}{2}
    \label{eq: breathing_parabola},
\end{equation}
where $\kappa$ represents the trapping (stiffness) constant, and $x_{c}$ is the position of the trap center. The values of the other relevant parameters in Eq. (\ref{eq: langevin}) are shown explicitly in Table~\ref{table_parameters}. These are realistic values commonly used in experimental setups, and the main results of this present work are obtained using these values, except where it is mentioned otherwise.

\begin{table}[ht] 
\caption{Parameters used in the numerical simulations. These are typical values in optical tweezers experiments with colloidal particles \cite{jones2015optical}.}
\centering
\begin{tabular}{c c c} 
\hline\hline 
Physical quantity & Representation & Value \\ [0.5ex] 
\hline
Particle's radius & $r$ & $\unit{1}{\micro \metre}$   \\ 
Particle's density & $\rho$ & $\unit{2.65}{\gram / \centi \metre^3}$   \\ 
Particle's mass & $m$ & $\unit{11}{\pico \gram}$   \\ 
Friction coefficient & $\gamma$ & $\unit{1.89 \cdot 10^{-8}}{\newton \second / \metre}$  \\ 
Medium's temperature & $T$ & $\unit{300}{\kelvin}$ \\ 
Initial trap stiffness & $\kappa$ & $\unit{1}{\pico \newton / \micro \metre}$\\
  [1ex] 
\hline
\end{tabular} 
\label{table_parameters} 
\end{table}

\section{Stochastic thermodynamics of driven Brownian motion \label{sec:stothermo}}

In the late 1990s, Sekimoto \cite{sekimoto1998langevin} showed that work and heat can be associated with individual trajectories of a Brownian particle. In our case, by varying the stiffness parameter $\kappa$ or the center position $x_c$ in time, the expression of the average work performed during the process is
\begin{equation}
    \langle W \rangle   = \int_{0}^{\tau} dt\,\frac{d \lambda (t)}{dt}\Biggr \langle \frac{\partial V(x,\lambda)}{\partial \lambda} \Biggr \rangle,
    \label{eq:average_work} 
\end{equation}
where $\lambda(t)$ denotes a control parameter varied according to a protocol of duration $\tau$. The symbol $\langle . \rangle$ denotes an average over many microscopic realizations with initial conditions sampled from an equilibrium distribution. 

In the previous expression for $\langle W\rangle$, the potential $V$ could be substituted by the Hamiltonian $H$ of the system, and the protocol could be written as
\begin{equation}
    \lambda (t) = \lambda_{i} + (\lambda_{f} - \lambda_{i})g(s),
    \label{eq:defg}
\end{equation}
where $s = t/\tau$, with boundary conditions: $g(0) = 0$, and $g(1) = 1$.

Here, we focus on the harmonic potential with time-dependent stiffness, $\lambda(t)=\kappa(t)$. Considering $x_{c} = 0$, we have the following time-dependent potential: 
\begin{equation}
    V(x,t) = \frac{\lambda(t)x^2}{2}.
    \label{eq:timeharmpoten}
\end{equation}

The values chosen for $\lambda_{i,f}$, corresponding to the initial and final values, are motivated by the experimental implementation in optical tweezers. For values too small, the particle may escape the detection region during the experiment, while for values too large, the variation of the particle's position may become difficult to detect.

Considering the time-independent version of potential (\ref{eq:timeharmpoten}), the Fokker-Planck equation (\ref{eq: Fokker-Planck}) can be solved to determine the dependence of the probability distribution with time \cite{riskenFPE,reichl1999modern}. For a given initial distribution, the solution has a characteristic time scale that roughly measures the relaxation time $\tau_{R}$ to the thermal distribution in the trap. A careful analysis shows that the relaxation time is $\tau_{R}=\gamma/(2\kappa)$ (see Sec.~\ref{sec:optslow} for the details) and using the values in Table \ref{table_parameters}, we find that it is equal to $\unit{9.4}{\milli \second}$. This natural time scale allows the classification of the protocols into slow or fast. Given a final trap stiffness, protocols with $\tau \ll \tau_R$ drive the system to regions far from equilibrium and are considered fast, while protocols with $\tau \gg \tau_R$ the system remains close to equilibrium throughout the process and are considered slow. 

As the process approaches the quasistatic limit, the average work tends to the Helmholtz free-energy difference, $\Delta F$, between the final and initial equilibrium states, in agreement with the Kelvin-Planck statement of the second law of thermodynamics \cite{callen1998thermodynamics},
\begin{equation}
      \langle W\rangle \geq \Delta F \,.
\end{equation}
The expression for $F(\lambda)$ can be obtained using statistical mechanics, and it is given by
\begin{eqnarray}
    F(\lambda) &=& - k_{B}T\ln Z \nonumber\\
    &=&  - k_{B}T\ln{\int \exp(-H(\Gamma,\lambda)/k_B T)d \Gamma}, 
    \label{eq:free-energy}
\end{eqnarray}
where
\begin{equation}
    Z = \int \exp(-H(\Gamma,\lambda)/k_B T)d \Gamma\,,
\end{equation}
is the partition function, $H(\Gamma,\lambda)=p^{2}/2m+V(x,\lambda)$ denotes the Hamiltonian of the Brownian particle and $\Gamma$ represents a point $(x,p)$ in phase space. For the harmonic potential with time-dependent stiffness, the free energy difference reads
\begin{equation}
    \Delta F(\lambda) = k_{B}T\ln{\left( \sqrt{ \frac{\lambda_{f}}{\lambda_{i}}} \right)},
    \label{eq:free-energy_har}
\end{equation}
where we used Eqs.~(\ref{eq: breathing_parabola}) and (\ref{eq:free-energy}), and $\lambda_{i}$ and $\lambda_{f}$ are the initial and final values of the stiffness parameter, respectively, as stated before.

\begin{figure}
  \begin{center}
  \centering
    \includegraphics[width=8cm]{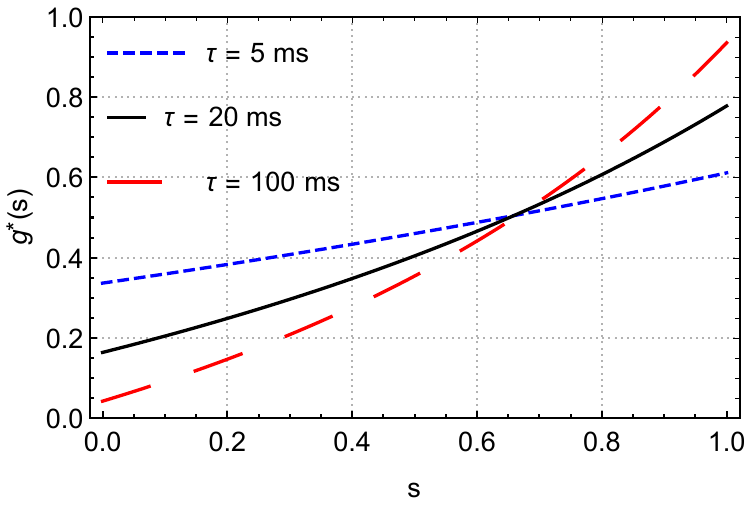}
  \end{center}
  \caption{Exact optimal protocol $\lambda^{*}(t)$ given by Eq.~(\ref{eq: optimal_exact}), and re-written as $\lambda^{*}(s)=\lambda_{i}+\Delta\lambda\,g^{*}(s)$, as a function of $s=t/\tau$ for different values of $\tau$. The initial and final values of $\lambda$ were chosen as $\lambda_{i} = \unit{1.0}{\pico \newton/\micro\meter}$ and $\lambda_{f} = \unit{2.5}{\pico \newton / \micro \meter}$, and $\Delta\lambda=\lambda_{f}-\lambda_{i}$. The faster the protocol is, the greater the initial and final jumps are.}
    \label{fig: protocol_exact}
\end{figure}

\section{Exact optimal protocols in the overdamped regime \label{sec:exactopt}}

In some cases, it is possible to derive exact expressions for the optimal protocols performed in driven overdamped Brownian motion. In this section, we review the main steps of the derivation since they will be our benchmark for analyzing the performance of optimal linear-response processes. Following Ref.~\cite{schmiedl2007optimal} and restricting ourselves to the change in the stiffness parameter, Eq.~(\ref{eq:average_work}) becomes
\begin{equation}
    \langle W \rangle  =  \frac{1}{2}\int_{0}^{\tau}dt\, \frac{d \lambda }{dt} \langle x^2   \rangle   = \frac{1}{2}\int_{0}^{\tau} dt\,\frac{d \lambda }{dt} w(t),
    \label{work_optimal1}
\end{equation}
where $w(t) = \langle x^2   \rangle$. The time evolution of  $w(t)$ can be obtained multiplying Eq.~(\ref{eq: Fokker-Planck}) by $x^{2}$ and integrating over $x$. This yields
\begin{equation}
     \frac{dw}{dt} =  - \frac{ 2\lambda}{\gamma} w  +  \frac{ 2 k_B T}{\gamma}.
     \label{time_evolution_w}
\end{equation}

This differential equation can be solved given an initial condition $w(0)$. Integrating Eq.~(\ref{work_optimal1}) by parts, we obtain
\begin{equation}
   \langle W \rangle =  \frac{1}{2}\left(\lambda(t) w(t) \Biggr|_{0}^{\tau} - \int_{0}^{\tau}  \lambda \frac{d w(t)}{dt}  dt \right).
   \label{eq: average_work3}
\end{equation}

Isolating $\lambda$ in Eq.~(\ref{time_evolution_w}) and substituting the result in the previous expression, we rewrite Eq.~(\ref{eq: average_work3}) as
\begin{multline}
\langle  W \rangle = \frac{1}{2} (w(t) \lambda(t) - k_{B}T \ln{w(t)})\Biggr|_{0}^{\tau} \\ + \frac{\gamma}{4} \int_{0}^{\tau} \frac{1}{w} \left(\frac{dw}{dt} \right)^2 dt. 
\end{multline}

To find the optimal protocol, one can first minimize the integral on the right-hand side. The minimization of this functional corresponds to solving the Euler-Lagrange equation
\begin{equation}
    \left(\frac{d w}{dt}\right)^2 - 2 w \frac{d^2 w}{dt^2} = 0,
\end{equation}
whose solution is
\begin{equation}
    w(t) = c_{1} (1 + c_{2}t)^2\,.
    \label{w_optimal}
\end{equation}

The $c_{1}$ and $c_{2}$ are constants to be determined. For instance, if initially, the particle is in thermal equilibrium, then $w(0) = c_{1} = k_{B} T/\lambda_{i}$. The other constant, $c_{2}$, we find by minimizing the work (\ref{eq: average_work3}) after using Eq.~(\ref{w_optimal}) with $c_{1}=k_{B}T/\lambda_{i}$,
\begin{equation}
    \langle W \rangle =  \frac{\lambda_{f}}{2 \lambda_{i}}(1 + \tau c_{2})^2 + \frac{(\tau c_{2})^2 \gamma}{ \lambda_{i} \tau}  
    - \frac{1}{2} - \ln(1+c_{2}\tau).
    \label{eq: exact}
\end{equation}

The value of $c_{2}$ that minimizes the previous expression is equal to 
\begin{equation}
      c_{2} \tau= \frac{- \gamma - \tau \lambda_{f} + \sqrt{\gamma^2 + 2\gamma \tau \lambda_{i} + \tau^2 \lambda_{f} \lambda_{i}}}{2 \gamma + \lambda_{f} \tau}.
      \label{c2}
\end{equation}

Finally, using Eqs.~(\ref{time_evolution_w}), (\ref{w_optimal}), and (\ref{c2}), we obtain the optimal protocol for the time-dependent stiffness,
\begin{equation} \label{eq: optimal_exact}
      \lambda^{*}(t) = \frac{\lambda_{i} - \gamma c_{2}(1 + c_{2} t)}{(1 + c_{2}t)^2}
\end{equation}

We remark that, for $t = 0$, the expression above leads to
\[
  \lambda^{*}(0) = \lambda_{i} - \gamma c_{2} \neq \lambda_{i}. 
\]
Thus, the optimal protocol $\lambda^{*}(t)$ has a discontinuity at the initial time $t=0$. The same happens for $t = \tau$, i.e., $\lambda^{*}(\tau) \neq \lambda_f$. Figure~\ref{fig: protocol_exact} illustrates these discontinuities showing that they decrease as the process becomes slower. 

\begin{figure}
  \begin{center}
  \centering
    \includegraphics[width=8cm]{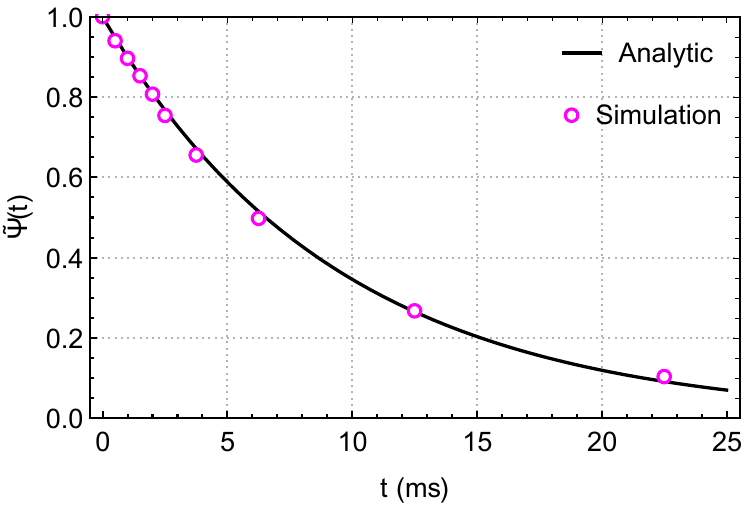}
  \end{center}
  \caption{ Comparison between analytical (Eq.~(\ref{eq: relaxation_breathing})) (black-solid line) and numerical calculations (magenta-dots) of $\tilde{\Psi}(t) = \Psi(t)/\Psi(0)$ (Eq.~(\ref{eq: relaxation_breathing0})) for the stiffening trap in the overdamped regime using the parameters values of Tabel~\ref{table_parameters}. Our numerical simulations integrate Eq.~(\ref{eq: langevin}) and allow for the numerical calculation of the averages in Eq.~(\ref{eq: relaxation_breathing0}). }
  \label{fig: relaxation_function}
\end{figure}

In addition, Fig.~\ref{fig: protocol_exact} shows that, for a fixed and not so small change $\Delta\lambda=\lambda_f-\lambda_i$ of the control parameter, larger values of $\tau$ lead to a more accentuated curvature of the exact optimal protocol. This implies that the rate of change $d\lambda^{*}/dt$ becomes stationary (apart from the jumps at the boundaries) as the process becomes faster. In other words, the solution (\ref{eq: optimal_exact}) contains the following physics: 
a time-dependent rate protocol is not a good optimization strategy if the protocol time $\tau$ becomes comparable with the relaxation time $\tau_{R}$. This will be corroborated by the perturbative approaches discussed in Secs.~\ref{sec:optslow} and \ref{sec:fastopt}.

\subsection{Limiting cases}
As verification of expression (\ref{eq: exact}), we can take the limits of arbitrarily short or long protocol duration. For extremely short protocols, we have
\begin{equation}
\lim_{\tau \to 0} c_{2}\tau  \to 0,
\end{equation}
and 
\begin{equation}
    \langle W \rangle \to \frac{1}{2}\left( \frac{\lambda_{f}}{\lambda_{i}} - 1\right),
\end{equation}
which is equal to the average work of the instantaneous protocol leading to a variation $\Delta\lambda=\lambda_f-\lambda_i$, as expected. For arbitrarily long protocols,
\begin{equation}
\lim_{\tau \to \infty} c_{2}\tau  \to  \left(\sqrt{\frac{\lambda_{f}}{\lambda_{i}}} - 1 \right),
\end{equation}
and
\begin{equation}
    \langle W \rangle \to  \Delta F = k_{B}T\ln{\left( \sqrt{ \frac{\lambda_{f}}{\lambda_{i}}} \right)}.
\end{equation}
So, the average work in the quasistatic limit is indeed equal to the free energy difference.

\subsection{Performance}
Equations~(\ref{eq: exact}) and (\ref{c2}) give the minimum average work required to change $\lambda(t)$ from $\lambda_{i}$ to $\lambda_{f}$ in a finite-time process of fixed duration $\tau$. In more complex cases, the work functional (\ref{eq:average_work}) may not be so easy to optimize, justifying the necessity of other optimization methods. To compare the perturbative approaches of Secs.~\ref{sec:optslow} and \ref{sec:fastopt} with the analytical solution presented in this section, we define the performance $\mathcal{P}$ as the relative difference between the average works $\langle W_{approx.} \rangle$ and $\langle W_{exact}\rangle$ performed along the approximate and the exact optimal protocols, respectively, i.e., 
\begin{equation}
    \mathcal{P} = \frac{\langle W_{approx} \rangle  - \langle W_{exact} \rangle }{\langle W_{exact} \rangle }.
    \label{eq: performance}
\end{equation}
We say that a method has a good performance when the value of the relative difference $\mathcal{P}$ is sufficiently small.

\begin{figure}
  \begin{center}
  \centering
    \includegraphics[width=8cm]{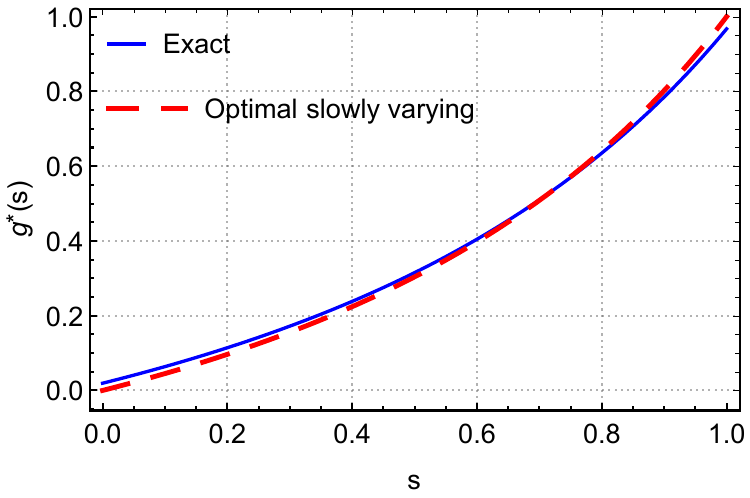}
  \end{center}
  \caption{ Comparison between the exact optimal protocol (\ref{eq: optimal_exact})(blue-solid line) and the optimal slowly-varying protocol (\ref{eq: optimal_close}) (red-dashed line) as a function of $s=t/\tau$. The relation between $\lambda(s)$ and $g(s)$ is given by Eq.~(\ref{eq:defg}). The parameters used are $\tau = \unit{200}{\milli \second}$, $\lambda_{i}=\unit{1.0}{\pico\newton/\micro\metre}$,  $\lambda_{f} = \unit{3.0}{\pico \newton / \micro \metre}$, and $\tau_{R}(\lambda_{i})=\unit{9.4}{\milli\second}$. The approximate optimal protocol (\ref{eq: optimal_close}) approaches the exact one as $\tau_{R}/\tau$ decreases.}
  \label{fig: protocol_slow}
\end{figure}

\section{Optimal slowly-varying processes \label{sec:optslow}}

The first perturbative approach we will discuss describes the work performed along slowly-varying processes. It is based on linear response theory (LRT) and on the assumption that, for slow enough processes, the relaxation to equilibrium happens much faster than the variation of the control parameter. In other words, we will deal with near-equilibrium processes in the vicinity of quasistatic variations. Next, we reproduce the main steps of deriving the functional measuring the energetic cost for this class of processes. We will follow closely Refs.~\cite{sivak2012prl,bonancca2014optimal,koning1997prb}. For an alternative derivation based on endoreversibility, see Ref.\cite{tsao1993jcp}.

In the linear response regime ($\Delta \lambda / \lambda_{i} \ll 1$), we can expand the Hamiltonian of the system of interest as
\begin{equation}
    H[\lambda(t)] = H(\lambda_{i}) + \Delta\lambda\, g(t)\partial_{\lambda} H + \mathcal{O}(\Delta \lambda^2),
\end{equation}
where we used Eq.~(\ref{eq:defg}) to express $\lambda(t)$ in terms of $g(t)$. Then, by using well-known methods of LRT \cite{kubo2012statistical}, we obtain the non-equilibrium average of the generalized force $\partial_{\lambda}H \equiv \partial H/\partial\lambda$,
\begin{multline}
\langle \partial_{\lambda} H(t) \rangle   = \langle \partial_{\lambda} H(0) \rangle_{eq; \lambda_{i}} + \chi_{0}^{\infty} \Delta \lambda g(t)\\- \Delta \lambda \int_{0}^{t}ds \phi(t-s)g(s), 
\label{eq: lr1}
\end{multline}
where $\langle \cdot \rangle_{eq; \lambda_{i}}$ is the average over the equilibrium canonical distribution, $\exp{(-H(\Gamma,\lambda)/k_{B}T)}/Z$, with control parameter $\lambda=\lambda_{i}$. The second term on the right-hand side describes the instantaneous response
\begin{equation}
    \chi_{0}^{\infty} = \bigg \langle \frac{\partial^2 H}{\partial \lambda^2} \bigg \rangle_{eq; \lambda_{i}},
\end{equation}
while the last term is the delayed response. The function $\phi(t)$ is the response function \cite{kubo2012statistical},
\begin{equation}
    \phi(t) = \langle \{ \partial_{\lambda} H(0), \partial_{\lambda} H(t)\} \rangle_{eq; \lambda_{i}} 
    \label{eq: respfunc}
\end{equation}
where $\{ \cdot ,\cdot \}$ is the Poisson bracket. Employing Kubo's formula, we find the relaxation function,  $\Psi(t)$,
\begin{equation}
    \Psi(t) = \beta \left(\langle \partial_{\lambda} H(0) \partial_{\lambda} H(t) \rangle_{eq;\lambda_{i}} - \langle \partial_{\lambda} H(0)\rangle_{eq;\lambda_{i}}^{2}\right)\,,
\label{eq: relaxation}
\end{equation}
 where $\phi(t) = - d \Psi(t)/dt$. Therefore, Eq.~(\ref{eq: lr1}) can be rewritten after an integration by parts as
\begin{multline}
\langle \partial_{\lambda} H (t) \rangle   = \langle \partial_{\lambda} H(0)\rangle_{eq;\lambda_{i}} - \tilde{\Psi} \Delta \lambda g(t)\\+ \Delta \lambda \int_{0}^{t}du\, \Psi(u)\frac{dg}{dt'} \biggr|_{t'=t-u}\,,
\label{eq:neqtforcelrt}
\end{multline}
where $\tilde{\Psi} = \Psi(0) - \chi_{0}^{\infty}$. 

Considering that $\Psi(t)$ decays sufficiently fast to assume $dg/dt$ is approximately constant within this time scale, the convolution in the right-hand side of the previous expression can be written as \cite{sivak2012prl,bonancca2014optimal}
\begin{eqnarray}
    \int_{0}^{t}du\, \Psi(u)\frac{dg}{dt'} \biggr|_{t'=t-u} &\approx& \frac{dg}{dt}\int_{0}^{t}du\,\Psi(u) \nonumber \\
    &\approx & \frac{dg}{dt}\int_{0}^{\infty}du\,\Psi(u) \,,
    \label{eq:timescaleapp}
\end{eqnarray}
where we have further assumed that extending the upper limit to infinity does not change the result significantly (this is probably justified only in cases where $\Psi(t)$ decays exponentially). 

Applying Eq.~(\ref{eq:neqtforcelrt}) and approximation (\ref{eq:timescaleapp}) to each infinitesimal variation of $\lambda$ along the protocol $\lambda(t)$, and plugging them in Eq.~(\ref{eq:average_work}) to compute the work performed, one finally finds the functional (see Refs.~\cite{sivak2012prl,bonancca2014optimal} for more details)
\begin{eqnarray}
    \langle W_{irr} \rangle &=& \langle W\rangle - \Delta F \nonumber\\
    &=&\frac{\beta (\Delta \lambda )^2}{\tau} \int_{0}^{1} ds \left(\frac{dg}{ds}\right)^2 \,\tau_{R}[g(s)]\,\chi [g(s)],
    \label{eq: slow}
\end{eqnarray}
for the so-called irreversible work $\langle W_{irr}\rangle$. The quantities $\tau_{R}[g(s)]$ and $\chi[g(s)]$ represent the parametric change of the relaxation time,
\begin{equation}
    \tau_{R}(\lambda) = \int_{0}^{\infty} \frac{\Psi(u)}{\Psi(0)} \,du \,,
    \label{eq: relax_time}
\end{equation}
and of the equilibrium fluctuations of the generalized force $\partial_{\lambda}H$,
\begin{equation}
    \chi(\lambda) = \Psi(0)/\beta = \langle ( \partial_{\lambda} H )^2 \rangle_{eq; \lambda} - \langle  \partial_{\lambda} H \rangle_{eq; \lambda}^2\,, 
    \label{eq: chi}
\end{equation}
along the protocol $\lambda(t)$.

Equation~(\ref{eq: slow}) provides the following physics: (i) it predicts a regime in which the energetic cost scales as $\tau^{-1}$ no matter the shape of the protocol, (ii) the protocol must be slowed down where the change of $\tau_{R} \cdot \chi$ \cite{note} increases in order to minimize the cost, and (iii) the power related to the irreversible loss of energy resembles Joule heating (it is proportional to $(dg/dt)^{2}$), and it is constant along optimal protocols with single control parameters \cite{sivak2012prl}.

The $\lambda^{*}(t)$ that minimizes Eq.~(\ref{eq: slow}) can be found via standard methods of calculus of variations once $\tau_{R}$ and $\chi$ are known. For the example of driven Brownian motion, exact analytical expressions can be obtained using Eqs.~(\ref{eq: relax_time}) and (\ref{eq: chi}) as we show next. For more complex situations, analytical approximations \cite{bonancca2014optimal} or numerical methods can be used \cite{koning2005jcp}.

\subsection{Relaxation function for the stiffening trap}

\label{ss: relaxation}

For the harmonic potential with time-dependent stiffness, the generalized force is simply
\begin{equation}
\partial_{\lambda}H = \frac{x^2}{2}.
\label{eq: genforce}
\end{equation}
Then, according to Eq.~(\ref{eq: relaxation}), the relaxation function in this case reads
 \begin{equation}
     \Psi(t) = \frac{\beta}{4} (\langle x(0)^2 x(t)^2\rangle_{eq; \lambda} - \langle x(0)^2 \rangle_{eq;\lambda}^2 ).
     \label{eq: relaxation_breathing0}
 \end{equation}

Hence, the relaxation function is found calculating the two-point equilibrium correlation of position squared. In other words, to obtain the relaxation function, we must measure/calculate the position, $x(t)$, and obtain the correlation $\langle x(0)^2 x(t)^2\rangle_{eq;\lambda}$ and the average $\langle x(0)^2 \rangle_{eq; \lambda}$ when the control parameter is kept fixed at the initial value. 
 
 \begin{figure}
  \begin{center}
  \centering
    \includegraphics[width=8cm]{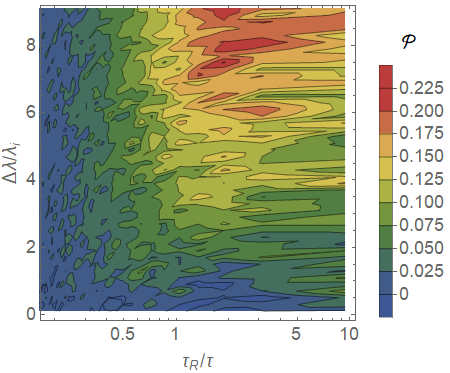}
  \end{center}
  \caption{ Performance $\mathcal{P}$, defined in Eq.~(\ref{eq: performance}), of the approximate optimal protocol (\ref{eq: optimal_close}). We recall that $\langle W_{exact}\rangle$ in Eq.~(\ref{eq: performance}) is obtained using Eq.~(\ref{eq: exact}). We have set $\lambda_{i} = \unit{1.0}{\pico \newton/ \micro \metre}$ and $\tau_{R}(\lambda_{i}) = \unit{9.4}{\milli \second}$. The other parameters were chosen according to Table~\ref{table_parameters}. The average work along (\ref{eq: optimal_close}) was obtained using $10^{5}$ trajectories.}
  \label{fig: map_slow}
\end{figure}
 
 To obtain an analytical expression, note that the trajectory of a particle in a harmonic potential can be found by solving the Langevin equation (\ref{eq: langevin}). The average values can be found using the properties of the thermal noise given by Eqs.~(\ref{Noise_mean_zero}) and (\ref{Noise_correlation}). Additionally, it is necessary to use the information that the system is initially in thermal equilibrium and that the initial conditions are distributed accordingly. Therefore, the relaxation function for the stiffening trap in the overdamped regime is equal to
\begin{equation}
    \Psi(t) =\Psi(0) \left( \frac{s_{2}e^{s_{1}|t|} - s_{1}e^{s_{2}|t|}}{s_{2} - s_{1}}\right)^2,
    \label{eq: relaxation_breathing}
\end{equation}
where
\begin{equation}
    \Psi(0) = \frac{1}{2\beta \lambda^2}, 
\end{equation}
\begin{equation}
    s_{1} = \frac{-\gamma/m - \sqrt{(\gamma/m)^2 - 4(\lambda/m)}}{2},
\end{equation}
and
\begin{equation}
    s_{2} = \frac{-\gamma/m + \sqrt{(\gamma/m)^2 - 4(\lambda/m)}}{2}.
    \label{eq: relaxff}
\end{equation}
Figure~\ref{fig: relaxation_function} compares the analytical expression (\ref{eq: relaxation_breathing}) with the relaxation function obtained from the numerical simulations of the Langevin equation (\ref{eq: langevin}).

\subsection{Optimal protocol for the stiffening trap}
In order to obtain the optimal protocol for the stiffening trap, we first use Eqs.~(\ref{eq: relaxation_breathing}) to (\ref{eq: relaxff}) in Eqs.~(\ref{eq: relax_time}) and (\ref{eq: chi}) to obtain expressions for $\tau_{R}$ and $\chi$. The latter reads
\begin{equation}
    \chi (\lambda) = \frac{1}{4} (\langle x(0)^4\rangle_{eq;\lambda} - \langle x(0)^2 \rangle_{eq;\lambda}^2 )= \frac{1}{2 (\beta \lambda)^2}\,,
\end{equation}
and the relaxation time is given by
\begin{equation}
    \tau_{R}(\lambda) = \frac{\gamma}{2 \lambda},
\end{equation}
when we consider the overdamped limit, i.e., after taking the limit $ m \to 0$ in Eq.~(\ref{eq: relaxation_breathing}).
Substituting the previous results in expression (\ref{eq: slow}) for $\langle W_{irr}\rangle$, we obtain
\begin{equation}
   \langle W_{irr} \rangle = \frac{ \gamma (\Delta \lambda)^2}{4\beta \tau \lambda_{i}^3} \int_{0}^{1} ds \left( \frac{dg}{ds}\right)^2 \frac{1}{ \left( 1 + \frac{\Delta \lambda}{\lambda_{i}}g(s) \right) ^3}.
   \label{eq: work_close_breathing}
\end{equation}

The minimum of this functional (after solving the Euler-Lagrange equation) is found for the protocol \cite{bonancca2014optimal}
\begin{equation}
    g^{*}(s) = -\frac{\lambda_{i}}{\Delta \lambda } + \frac{1}{A(s+B)^2},
    \label{eq: optimal_close}
\end{equation}
where $A$ and $B$ are given by the boundary conditions $g^{*}(0) = 0$ and $g^{*}(1) = 1$. The protocol (\ref{eq: optimal_close}) is depicted in Fig.~\ref{fig: protocol_slow}, where we see how well it approximates the exact optimal protocol (\ref{eq: optimal_exact}) for a small value of $\tau_{R}/\tau$. 

The numerical analysis of the performance of this optimal protocol is part of our primary goal. This is shown in Fig.~\ref{fig: map_slow}, where the coefficient $\mathcal{P}$ (Eq.~(\ref{eq: performance})) is obtained for an extensive variation of the relative change, $\Delta\lambda/\lambda_{i}$, of the control parameter and of the protocol duration $\tau$. We remind that the work performed along the exact optimal protocol (\ref{eq: optimal_exact}), denoted in Eq.~(\ref{eq: performance}) by $\langle W_{exact}\rangle$, is given by Eq.~(\ref{eq: exact}). The quantity $\langle W_{approx}\rangle $ used in $\mathcal{P}$ is obtained from numerical simulations of Eq.~(\ref{eq: langevin}) using a time-dependent stiffness parameter given by Eq.~(\ref{eq: optimal_close}). The numerical average leading to $\langle W_{approx}\rangle$ was obtained using $10^{5}$ microscopic realizations.

As expected, the optimal protocol (\ref{eq: optimal_close}) performs better in the regime of slow processes, i.e., when $\tau_{R}/\tau < 1$. Increasing both $\Delta\lambda/\lambda_{i}$ and $\tau_{R}/\tau$, we enter the far-from-equilibrium region, and the performance of (\ref{eq: optimal_close}) decreases substantially. The irregular boundaries between regions with different values of $\mathcal{P}$ is a consequence of our finite statistical sample. We remark that Fig.~\ref{fig: map_slow} does not check the range of validity of the functional (\ref{eq: work_close_breathing}). Instead, it tells us how well the optimal protocol obtained from this functional performs when compared with the protocol obtained from the exact solution (\ref{eq: optimal_exact}).

\begin{figure}
  \begin{center}
  \centering
    \includegraphics[width=8cm]{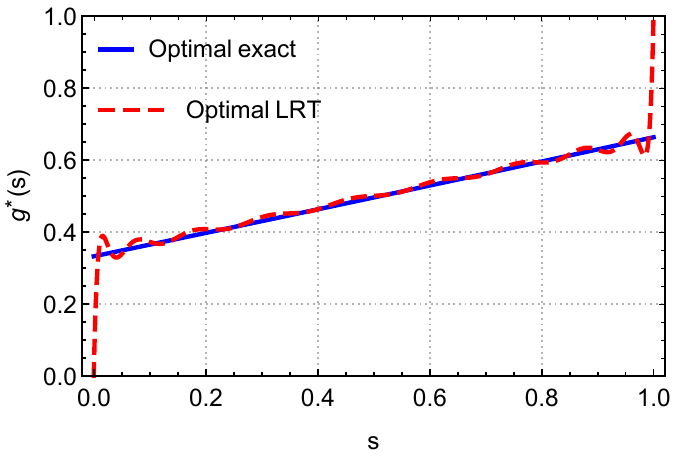}
  \end{center}
  \caption{  Comparison between the exact optimal protocol (\ref{eq: optimal_exact}) (blue-solid line) and the protocol that minimizes (\ref{eq: irr_work}) (red-dashed line) as a function of $s=t/\tau$. The relation between $\lambda(s)$ and $g(s)$ is given by Eq.~(\ref{eq:defg}). These results refer to $N=10$, $\lambda_{i}=\unit{1.0}{\pico \newton / \micro \metre}$, $\lambda_{f} = \unit{1.1}{\pico \newton / \micro \metre}$, $\tau = \unit{9}{\milli \second}$, and $ \tau_{R} = \unit{9.4}{\milli \second}$. We used the relaxation function (\ref{eq: relaxation_breathing}) with the parameters values in Table~\ref{table_parameters}.}
  \label{fig: protocol_weak}%
\end{figure}

\section{Fast but weak optimal processes \label{sec:fastopt}}

Linear response theory can also be used to describe a regime that is somewhat complementary to the one described previously. In other words, we will use LRT to describe a regime in which the protocols can be arbitrarily fast but restricted to small variations of the control parameter, i.e., $\Delta\lambda/\lambda_{i} < 1$. The functional for $\langle W_{irr}\rangle$ can be obtained in this case using, again, the nonequilibrium average (\ref{eq:neqtforcelrt}) of the generalized force. However, we will keep the convolution in it, that is, we will not perform the approximation (\ref{eq:timescaleapp}) since there will be no clear time-scale separation between the change in $\lambda(t)$ and the relaxation to equilibrium.

Reference~\cite{acconcia2015degenerate} shows that by plugging Eq.~(\ref{eq:neqtforcelrt}) in Eq.~(\ref{eq:average_work}) leads to
\begin{multline}
 \langle W_{irr} \rangle = \\\frac{(\Delta \lambda )^2}{2} \int_{0}^{1}ds \int_{0}^{1}ds' \Psi(\tau (s-s'))\frac{d{g}(s)}{ds}\frac{dg(s')}{ds'}\,,
    \label{eq: irr_work}
\end{multline}
where $s=t/\tau$, and $\Psi(t)$ is given again by Eq.~(\ref{eq: relaxation}). An effective strategy to minimize (\ref{eq: irr_work}) is that described in Ref.~\cite{bonancca2018minimal}. It consists of expanding $dg/ds$ in some basis of functions in the interval $s \in [0,1]$. Due to its convenient mathematical properties, a good choice of basis is that formed by the Chebyshev polynomials $T_{n}(x)$ \cite{weisse2006kernel}. Following Refs.~\cite{bonancca2018minimal,weisse2006kernel}, the truncated and regularized expansion of $dg/ds$ in a finite number $N$ of polynomials $T_{n}(x)$ in the interval $[0,1]$, then, reads 
\begin{equation}
    \frac{dg(s)}{ds} = \sum_{n=1}^{N} a_{n}\,g_{N,n}\,T_{n}(2s-1),
    \label{eq: protocol_expansion}
\end{equation}
where $a_{n}$ are the coefficients to be determined and the factors $g_{N,n}$ regularize the truncated expansion \cite{weisse2006kernel} to avoid the Gibbs phenomenon at the extremities of the expansion interval. Their expression is \cite{weisse2006kernel}
\begin{multline}
    g_{N,n} = \frac{N-n+1}{N+1}\cos{\left(\frac{\pi n}{N+1}\right)} \\+\frac{1}{N+1} \sin{\left(\frac{\pi n}{N+1} \right)} \cot \left( \frac{\pi}{N+1}\right) .
\end{multline}

\begin{figure}
  \begin{center}
  \centering
    \includegraphics[width=8cm]{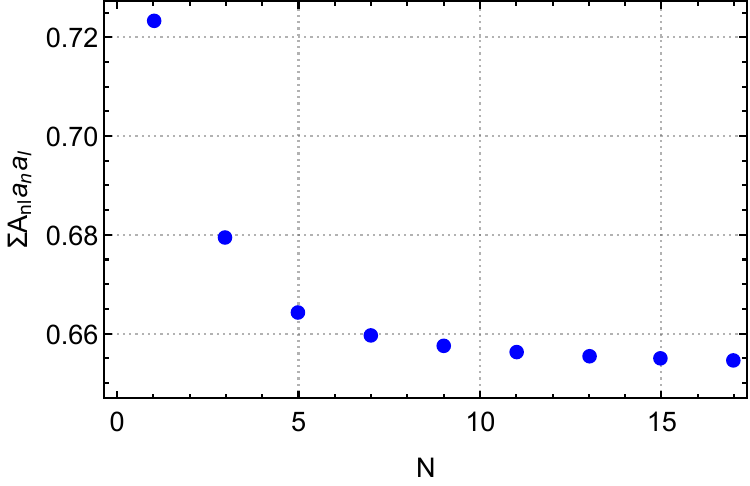}
  \end{center}
  \caption{Numerical evaluation of the right-hand side of Eq.~(\ref{eq: irr_work2}), $\sum_{n,l}^{N} A_{nl}a_{n}a_{l}$, for different values of $N$, i.e., the number of polynomials used in (\ref{eq: protocol_expansion}). We used the coefficients $a_{n}$ corresponding to the optimal protocol. The factors $A_{nl}$ were calculated using expression (\ref{eq: relaxation_breathing}) for $\Psi(t)$ and the parameters values in Table~\ref{table_parameters}. In particular, $\lambda_{i}=\unit{1.0}{\pico\newton/\micro\metre}$ and $\tau=\unit{9.0}{\milli\second}$.}
  \label{fig: convergence}
\end{figure}

Substituting expression (\ref{eq: protocol_expansion}) into Eq.~(\ref{eq: irr_work}), we obtain
\begin{equation}
    \langle W_{irr} \rangle [(\Delta \lambda )^2 \Psi(0)/2]^{-1} = \sum_{n,l}^{N} A_{nl}a_{n}a_{l}\,,
    \label{eq: irr_work2}
\end{equation}
where the $A_{nl}$ are given by
\begin{multline}
 A_{nl} \\= \int_{0}^{1} \int_{0}^{1} \tilde{\Psi}\left(\tau (s-s')\right) \, g_{N,n} g_{N,l} T_{n}(2s-1) T_{l}(2{s}'-1)ds' ds,
\end{multline}
with $\tilde{\Psi}(t) = \Psi(t)/\Psi(0)$. The irreversible work (\ref{eq: irr_work}) becomes a finite multidimensional quadratic form whose minimum we want to find. The coefficients $a_{n}$ that give such minimum value also have to obey the boundary conditions $g(0) = 0$ and $g(1) = 1$, which work as additional constraints in our optimization. Using the method of Lagrange multipliers, we can find the coefficients $a_{n}$ that provide the optimal protocol by solving a set of linear algebraic equations.

We remark that the relaxation function $\Psi(t)$ is the main physical input to the optimization problem. The factors $A_{nl}$ crucially depend on the protocol time $\tau$ and $\Psi(t)$. Due to its relation with the response function (\ref{eq: respfunc}), $\Psi(t)$ can be obtained from experiments when it is not accessible theoretically. This is an advantage of this approximate method, which can be readily applied to general potentials for which one does not have an exact solution for the optimal protocol. 

Figure~\ref{fig: protocol_weak} shows the optimal protocol obtained from minimizing (\ref{eq: irr_work2}) for specific values of $\tau$ and $N$. Although approaching the exact optimal solution (\ref{eq: exact}) in most part of the interval, the approximate optimal protocol clearly has smooth versions of the jumps presented by expression (\ref{eq: exact}). As shown in Ref.~\cite{bonancca2018minimal}, such smooth jumps decrease as the value of $\tau$ increases and the process becomes slower, in agreement with the behavior of the exact solution (\ref{eq: exact}). Reference~\cite{bonancca2018minimal} also shows that these smooth but steep features become sharper as the number of polynomials in (\ref{eq: protocol_expansion}) increases, providing a better agreement with the exact solution. However, Fig.~\ref{fig: convergence} shows that the value of $\langle W_{irr}\rangle$ given by (\ref{eq: irr_work}) does not change considerably for $N \gtrsim 10$. In other words, the optimization method just described is already efficient for low values of $N$.

\begin{figure}
  \begin{center}
  \centering
    \includegraphics[width=8.5cm]{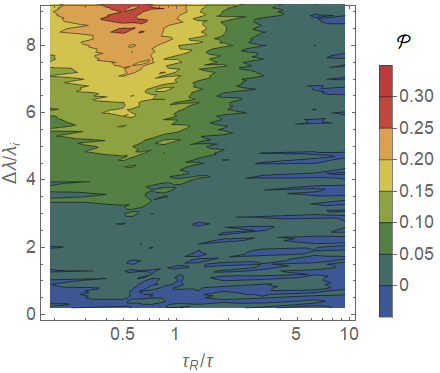}
  \end{center}
  \caption{  Performance $\mathcal{P}$, defined in Eq.~(\ref{eq: performance}), of the optimal protocol that minimizes (\ref{eq: irr_work}) and obtained from the methods of Sec.~\ref{sec:fastopt} for $N=10$. We recall that $\langle W_{exact}\rangle$ in Eq.~(\ref{eq: performance}) is obtained using the exact optimal protocol (\ref{eq: exact}) and Eq.~(\ref{work_optimal1}). We have set $\lambda_{i} = \unit{1.0}{\pico \newton/ \micro \metre}$ and $\tau_{R}(\lambda_{i}) = \unit{9.4}{\milli \second}$. The other parameters were chosen according to Table~\ref{table_parameters}. The average work along the approximate optimal protocol was obtained from numerical simulations of Eq.~(\ref{eq: langevin}) using $10^{5}$ trajectories.}
  \label{fig: map_weak}
\end{figure}

Figure~\ref{fig: map_weak} shows the performance $\mathcal{P}$ (Eq.~(\ref{eq: performance})) of the approximate optimal protocols that minimize (\ref{eq: irr_work}) for an extensive variation of the relative change $\Delta\lambda/\lambda_{i}$ of the control parameter, and of the protocol duration $\tau$. Analogously to the results of Fig.~\ref{fig: map_slow}, the work performed along the approximate optimal protocol was obtained from the numerical simulations of Eq.~(\ref{eq: langevin}) using $10^{5}$ trajectories. Again, the performance is excellent in the region of $\Delta\lambda/\lambda_{i}\ll 1$. Nevertheless, it remains very good in most part of the far-from-equilibrium region. This is in clear contrast to the performance of optimal protocols obtained in Sec.~\ref{sec:optslow}, using the geometric approach. 

\begin{figure}
  \begin{center}
  \centering
    \includegraphics[width=8.0cm]{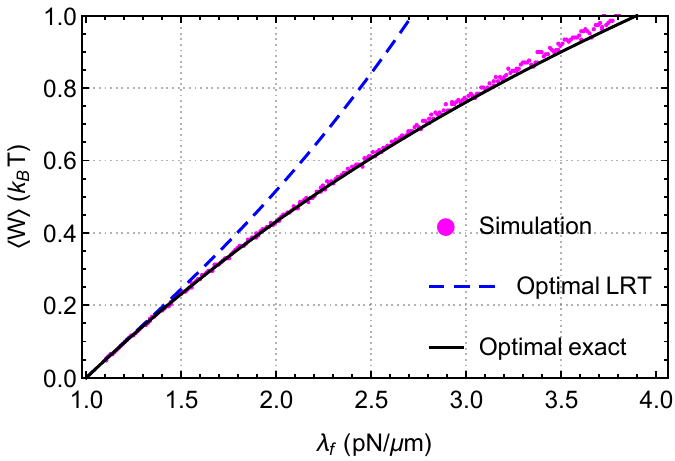}
  \end{center}
  \caption{ Average work (\ref{eq:average_work}) as a function of $\lambda_{f}$ for $\tau = \unit{9.0}{\milli \second}$ and $\lambda_{i} = \unit{1.0}{\pico \newton/ \micro \metre}$. The other parameters were chosen according to Table~\ref{table_parameters}. The black-solid line shows Eq.~(\ref{eq: exact}). The blue-dashed line shows Eq.~(\ref{eq: irr_work}) for its optimal protocol with $N=10$. Using this protocol, we have simulated the Langevin dynamics (\ref{eq: langevin}) and obtained the average work shown in dots and magenta using $10^{5}$ trajectories. The good performance of the approximate optimal protocol goes beyond the range of validity of expression (\ref{eq: irr_work}).}
  \label{fig: average_work_weak}%
\end{figure}

The outstanding performance of the protocols minimizing (\ref{eq: irr_work}) far from equilibrium is substantiated in Fig.~\ref{fig: average_work_weak}. There, it is shown that when we drive the Brownian particle using the approximate optimal protocols, the work performed is almost indistinguishable from that using the exact optimal protocol (\ref{eq: exact}), even when $\Delta\lambda/\lambda_{i} > 1$. The validity of Eq.~(\ref{eq: irr_work}) is, however, restricted to a small range.

\begin{figure}
  \begin{center}
  \centering
    \includegraphics[width=8cm]{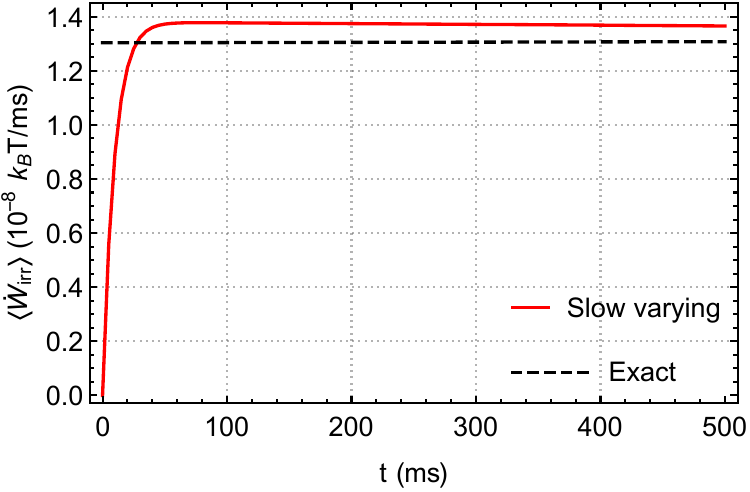}
  \end{center}
  \caption{ Comparison between the excess power (\ref{eq: excpower}) along the exact optimal protocol (\ref{eq: optimal_exact}) (black-dashed line) and the optimal slowly-varying protocol (\ref{eq: optimal_close}) (red-solid line). The parameters are $\tau = \unit{500}{\milli \second}$, $\tau_{R}=\unit{9.4}{\milli\second}$, $\lambda_{i}=\unit{1.0}{\pico \newton / \micro \metre}$, and $\lambda_{f} = \unit{3.0}{\pico \newton / \micro \metre} $.}
  \label{fig: power1}%
\end{figure}

\section{Excess power \label{sec:power}}

\begin{figure}
  \begin{center}
  \centering
    \includegraphics[width=8cm]{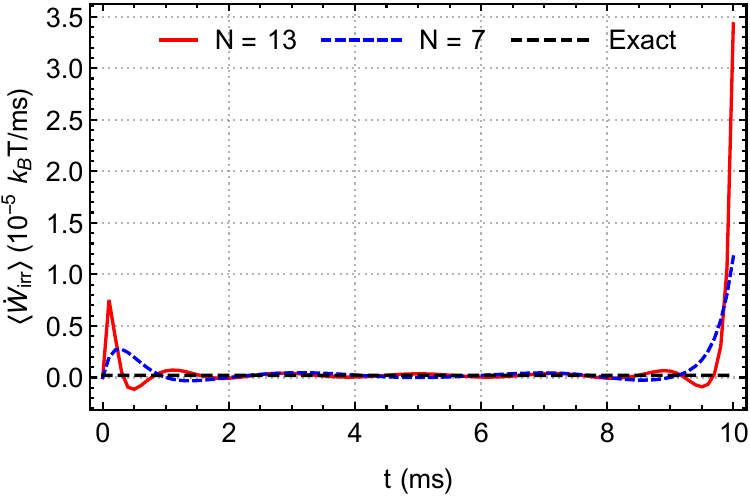}
  \end{center}
  \caption{  Excess power (\ref{eq: excpower}) for different optimal protocols using parameters $\tau = \unit{10}{\milli \second}$, $\tau_{R}=\unit{9.4}{\milli \second}$, $\lambda{i}=\unit{1.0}{\pico \newton / \micro \metre}$, and $\lambda_{f} = \unit{1.2}{\pico \newton / \micro \metre} $. The red-solid line and the blue-dashed line depict the excess power along the protocols minimizing (\ref{eq: irr_work}) with $N = 13$ and $N = 7$, respectively. The black-dashed line depicts the excess power along the exact optimal protocol (\ref{eq: optimal_exact}).}
  \label{fig: power2}
\end{figure}

\begin{figure}
  \begin{center}
  \centering
    \includegraphics[width=8cm]{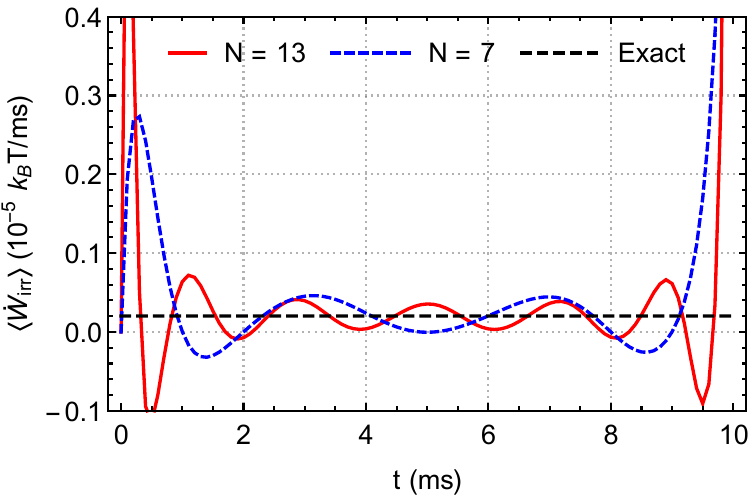}
  \end{center}
  \caption{ Zoom-in of Fig.~\ref{fig: power2}. The excess power (\ref{eq: excpower}) along the approximate optimal protocols of Sec.~\ref{sec:fastopt} oscillates around the corresponding value of excess power along the exact optimal protocol (\ref{eq: optimal_exact}). As $N$ increases, the oscillations decrease in the central part of the protocol. }
  \label{fig: power3}
\end{figure}

Figures~\ref{fig: power1} to \ref{fig: power3} show the comparison between the exact excess power, i.e., the excess power along the exact optimal protocol (\ref{eq: optimal_exact}), and that obtained from the approximate optimal protocols of Secs.~\ref{sec:optslow} and \ref{sec:fastopt}. By excess power, we mean the quantity whose integral gives the irreversible work $\langle W_{irr}\rangle = \langle W\rangle - \Delta F$. For instance, Eq.~(\ref{eq:average_work}) gives the total work as the integral of the total power. Thus, the excess power does not account for the power delivered to change the free energy. Mathematically, this can be translated as
\begin{equation}
 \langle \dot{W}_{irr}\rangle = \frac{d\lambda}{dt}\left( \left\langle \frac{\partial H}{\partial\lambda}\right\rangle - \frac{\partial F}{\partial\lambda}\right)\,,
 \label{eq: excpower}
\end{equation}
where $\partial F/\partial\lambda$ denotes the derivative of the free energy. Both $\langle \partial_{\lambda}H\rangle = \langle x^{2}\rangle/2$ and $\partial_{\lambda}F$ must be evaluated along the protocol $\lambda(t)$.

Reference~\cite{sivak2012prl} shows that the functional~(\ref{eq: slow}) for a single control parameter predicts a constant excess power along the optimal protocol. Although this reveals an important feature of the physics of slowly-varying optimal protocols, it remains to be verified whether this also applies to other non-equilibrium regimes. Figure~\ref{fig: power1} compares the excess power (\ref{eq: excpower}) obtained from the numerical simulations of Eq.~(\ref{eq: langevin}) using the optimal protocol (\ref{eq: optimal_close}) and the exact optimal protocol (\ref{eq: optimal_exact}) for $\tau_{R}/\tau < 1$. The initial variation is expected since in the exact dynamics there is no driving exactly before the process begins. 

Figures~\ref{fig: power2} and \ref{fig: power3} compare the numerical calculations of the excess power using the approximate optimal protocol that minimizes (\ref{eq: irr_work}) in the regime of fast-but-weak processes. The exact optimal protocol leads once more to a constant value, whereas the approximate one yields oscillates around this value. As $N$ increases, our results point to a decrease of the oscillations taking place in the central part of the protocol, becoming, however, more salient at the extremities.

\section{Conclusion \label{sec:conclusion}}

Using an overdamped driven Brownian particle as a benchmark, we compared the performance of two classes of approximate optimal protocols to the exact optimal solution,  performing numerical simulations with realistic parameters and presenting it in units and scales relevant to current experiments. 
Generally, the approximate optimal protocols have excellent performances in the regions where they were expected. However, one of our main results is how effectively good the performance can be even far outside the region where the approximation is expected to be valid. Hence, our results help to determine the range of validity of each of the perturbative approaches. Moreover, due to the difficulties controlling the approximations involved in the linear-response descriptions (see Refs.~\cite{blaber2020jcp,naze2022jsp}), this numerical determination is a welcome achievement in itself.

Our analysis shows a clear advantage of the perturbative formulation in describing the optimal energetic cost far from equilibrium compared to the geometric approach. In particular, we verified that the performance of protocols derived from the geometric methods decreases considerably as the duration of the process becomes comparable to the relaxation time.  

Despite underperforming in the region of slowly-varying processes, the linear response method for fast but weak optimal protocols performs exceptionally well in other regions. Furthermore, these approximate optimal protocols can be more easily implemented experimentally due to their smooth character.

In this case, the excess power along them has clear oscillations and can even show negative values. This seems to point out that non-monotonic driving might do a better job far from equilibrium as long as it combines oscillations with a steady change, as shown in our results for the excess power. This fact supports the claim that the perturbative approaches may increase our physical understanding of optimal nonequilibrium processes compared to purely numerical optimization methods.

\section*{Acknowledgements \label{sec:acknowledgements}}

%\emph{Acknowledgements --}
LPK acknowledges financial support by CNPq (Centro Nacional de Desenvolvimento Científico e Tecnológico), Grant n.o 131013/2020-3. MVSB and SRM acknowledge financial support by FAPESP (Funda\c{c}\~ao de Amparo \`a Pesquisa do Estado de S\~ao Paulo), Grants n.o 2020/02170-4 and 2019/27471-0.

\bibliography{Ref.bib}

\end{document}